# OBSERVATIONS OF THE CATACLYSMIC VARIABLE SDSS J081321.91+452809.4

JEREMY SHEARS [1], IAN MILLER [2], STEVE BRADY [3]

1) British Astronomical Association (BAA) Variable Star Section, "Pemberton", School Lane, Bunbury, Tarporley, Cheshire, CW6 9NR, UK, bunburyobservatory@hotmail.com
2) British Astronomical Association (BAA) Variable Star Section, Furzehill House, Ilston, Swansea, SA2 7LE, UK, furzehillobservatory@hotmail.com
3) American Association of Variable Star Observers (AAVSO), 5 Melba Drive, Hudson, NH 03051, USA, sbrady10@verizon.net

**Abstract:** Our observations of the first reported outburst of SDSS J081321.91+452809.4 during 2008 April show that this cataclysmic variable is a dwarf nova. The outburst amplitude was at least 3.1 magnitudes and the outburst appears to have been rather short-lived at around 3 days with a rapid decline to quiescence of 0.73 mag/day.

## Introduction

Cataclysmic variables stars (CVs) are semi-detached binary systems in which a white dwarf accretes matter from a cool main sequence star via an accretion disc. CVs include novae, dwarf novae and nova-like variables which are distinguished by their amplitudes and timescale of variability.

SDSS J081321.91+452809.4 (hereafter "SDSS 0813") was identified as a CV during the Sloan Digital Sky Survey (SDSS). Szkody *et al. (*2002) presented a spectrum in which the presence of a late-type companion was noted, suggesting that it may have a long orbital period. A spectroscopic study by Thorstensen et al. (2004) yielded an orbital period of 0.2890(4) d, although a less favoured period of 0.2867(4) d could not be ruled out. They commented that SDSS 0813 is notable as a long orbital period CV which is not known to erupt, although outbursts might have been overlooked. Based on a quiescence magnitude of V=18.4, they suggested that it may reach V = 14.7 at outburst.

## The 2008 April outburst of SDSS 0813

SDSS 0813 was detected in outburst for the first time on 2008 April 6.952 by Miller (2008) as part of a campaign by the authors to monitor this CV for possible outbursts. Time resolved unfiltered (C = clear) photometry was conducted during the outburst using the instrumentation shown in Table 1 according to the log shown in Table 2. In all cases raw images were flat-fielded and dark-subtracted, before being analysed using the ensemble aperture photometry software function in AIP4WIN version 2 (Berry and Burnell 2005). We adopted an ensemble of the comparison stars 132, 140, 147 and 152 in the chart for SDSS 0813 by Simonsen (2002) dated 021022 and the V magnitude sequence of Henden & Simonsen (2002). Typically the error on the measurements was 0.02 mag when the star was at ~15.5C, 0.03 mag at ~16.0C and 0.05 mag when the star had faded to between 16.5-17.0C.

The light curve of the outburst is shown in Figure 1. At its brightest it reached magnitude 15.3C implying an outburst amplitude of at least 3.1 mag. However, the onset of the outburst is not well defined and the object was already in decline at discovery, which suggests that the outburst was already well underway when it was discovered. The observation preceding the outburst detection was 7 days earlier (JD 2454556.62794 at 18.1C). After 3 days the star had already declined to 17.6C, on the way back to quiescence, showing the outburst to be rather short. The average decline rate over this period was a rapid 0.73 mag/d and no modulations were detected. Photometry from the longer time series runs is shown on an expanded scale for clarity in Figures 2 and 3.

| **Observer** | **Telescope** | **CCD** |
|---|---|---|
| JS | 0.1 m fluorite refractor | Starlight Xpress SXV-M7 |
| SB | 0.4 m reflector | SBIG ST-8XME |
| IM | 0.35 m SCT | Starlight Xpress SXVF-H16 |

Table 1: Equipment used

| Date in 2008 (UT) | Start JD | Duration (h) | Integration time (s) | Observer |
|---|---|---|---|---|
| April 6 | 2454563.45049 | 0.6 | 30 | IM |
| April 7 | 2454564.35432 | 5.5 | 30 | IM |
| April 7 | 2454564.36734 | <0.1 | 60 | JS |
| April 8 | 2454565.37770 | 3.6 | 30 | IM |
| April 9 | 2454565.53293 | 4.9 | 120 | SB |
| April 9 | 2454566.35565 | 0.2 | 30 | IM |
| April 9 | 2454566.37215 | <0.1 | 60 | JS |
| April 13 | 2454570.36397 | 0.2 | 30 | IM |

Table 2: Log of time-series observations

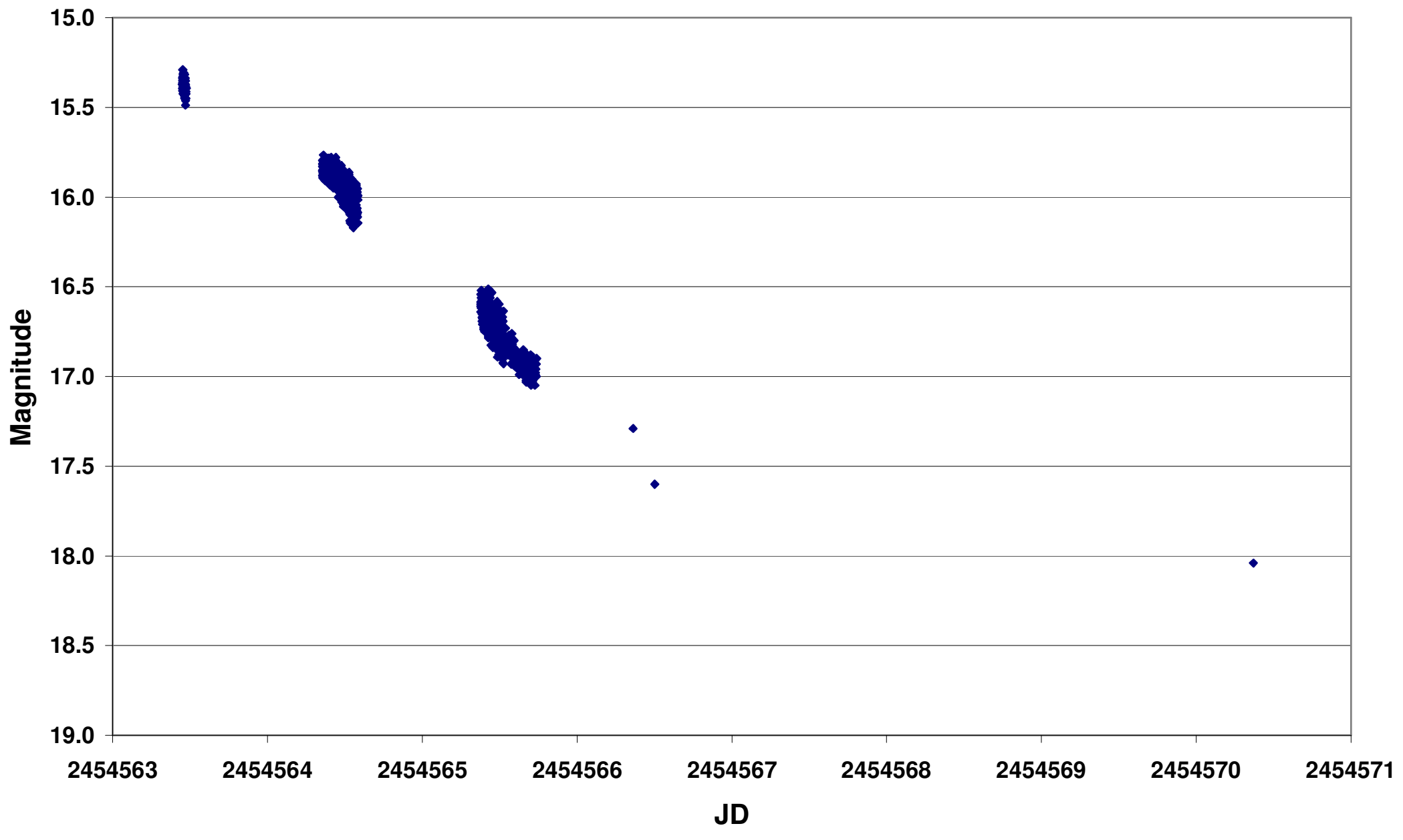

Figure 1: Light curve of the 2008 April outburst

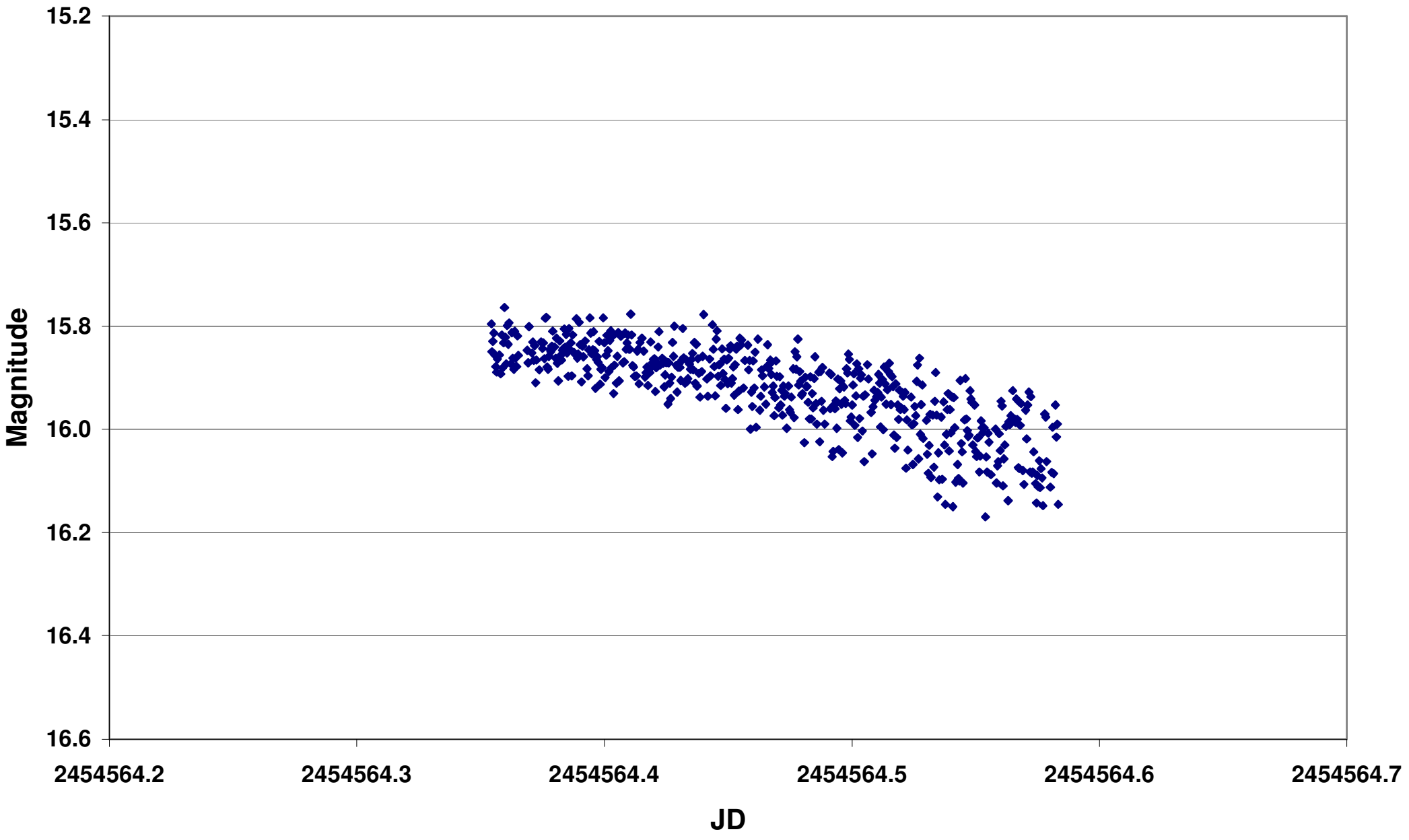

Figure 2: Time series photometry on JD 2454564

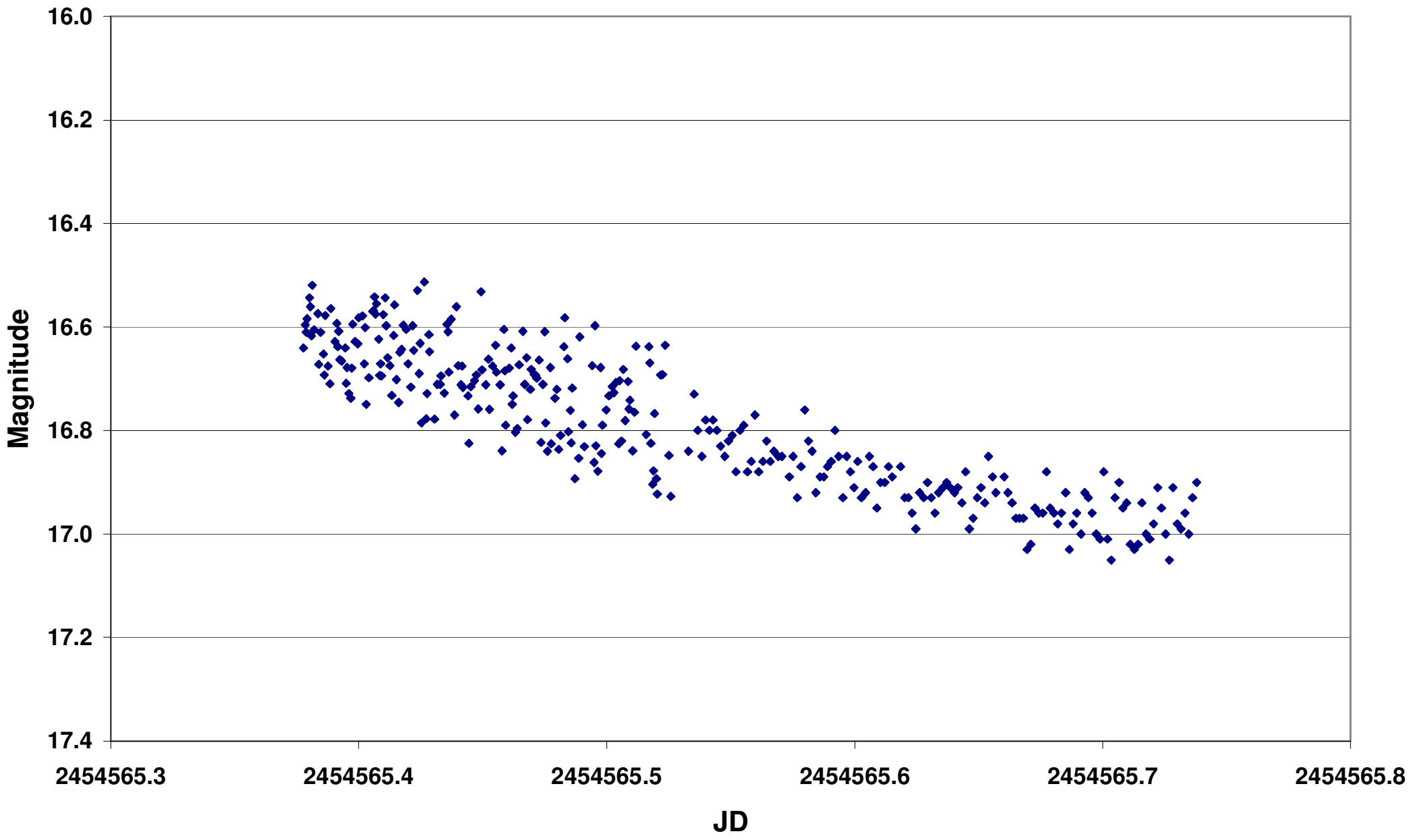

Figure 3: Time series photometry on JD 2454564

## Historical data on SDSS 0813

Figure 4 shows the long term light curve for SDSS 0813 based on data from the AAVSO International Database, supplemented by data from the authors, covering the period JD 2452310.7 to 2454607.4 (2002 February 5 to 2008 May 20), which clearly shows the 2008 April outburst. The frequency of observations of SDSS 0813 increased from late 2007 onwards as the star was added to the Recurrent Objects Programme, a joint project between the British Astronomical Association and *The Astronomer* magazine to encourage observations of poorly characterised eruptive variable stars (Poyner 2008).

However there are large gaps in the observational record which makes it difficult to draw conclusions about the outburst frequency. Furthermore, it is possible that further outbursts have been missed during this period due to the incomplete observational coverage and a low accuracy of the data. This is especially if the rather short duration of the 2008 April outburst is typical of outbursts in SDSS 0813. We note that there are several occasions when the star was as bright as magnitude 17.5+/0.06, i.e. ~ 1 mag above normal quiescence. Whether these simply represent outbursts that have been missed due to incomplete coverage, or activity at quiescence, or whether they are brief attenuated outbursts as seen in other dwarf novae such as V1316 Cyg (Shears *et al*. 2006), is unclear at the moment. However, we cannot draw definite conclusions from quiescence measurements due to the rather large errors on the measurements at quiescence, typically 0.07 to 0.09 mag, which contribute to the apparent scatter in the data points at quiescence.

**Conclusion**

Our observations of the first reported outburst of SDSS 0813 show that this star is a dwarf nova. The outburst amplitude was at least 3.1 magnitudes and the outburst appears to have been rather short at around 3 days with a rapid decline to quiescence of 0.73 mag/d. The outburst may actually have been longer and with a greater amplitude, but it is likely that the first part of the outburst was missed. The fact that SDSS 0813 is a dwarf nova having a relatively long orbital period, which places it above the so-called period gap in the distribution of the orbital periods of dwarf novae, indicates that it is likely a member of the SS Cyg family (GCVS classification "UGSS").

Further observations of SDSS 0813 are encouraged to investigate the outburst frequency, outburst amplitude and whether the dwarf nova undergoes frequent brief outbursts (see Poyner 2008).

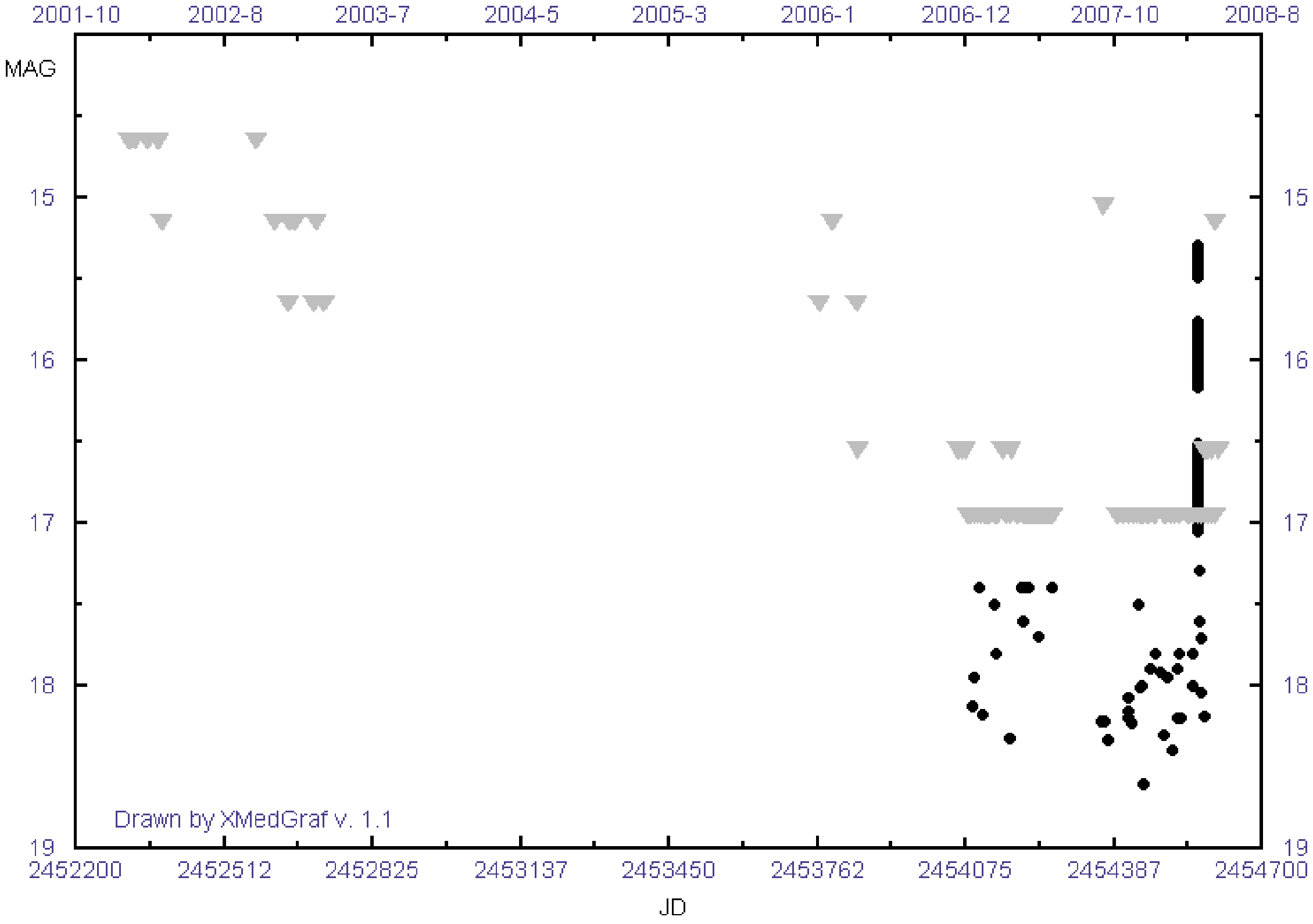

Figure 4: Long term light curve
Positive detections are denoted by a circle, whereas negative observations ("fainter thans") are denoted by a triangle. Data are from the authors and the AAVSO International Database.

**Acknowledgements**

The authors gratefully acknowledge the use of observations from the AAVSO International Database contributed by observers worldwide, the use of SIMBAD, operated through the Centre de Données Astronomiques (Strasbourg, France), and the NASA/Smithsonian Astrophysics Data System. We thank Gary Poyner, co-ordinator of the Recurrent Objects Programme, for his encouragement and Dr. Boris Gaensicke (University of Warwick, UK) for helpful comments during the preparation of this paper. We are indebted to Mike Simonsen for producing the chart for SDSS 0813 and making it available on his web site. We thank Lubos Brát for preparing Figure 4.